\documentclass[a4paper]{article}

\usepackage{amssymb}
\usepackage{latexsym}
\usepackage{amsmath}
\usepackage{cite}

\newcommand{\Mat}{\,{\rm Mat}\,}

\newcommand{\mbold}[1]{\mbox{\boldmath{\ensuremath{#1}}}}

\begin{document}
\title{Homogeneous Plane-wave Spacetimes and their Stability}
\author{Sigbj{\o}rn Hervik\thanks{S.Hervik@damtp.cam.ac.uk}, Hari K. Kunduri\thanks{H.K.Kunduri@damtp.cam.ac.uk}, James Lucietti\thanks{J.Lucietti@damtp.cam.ac.uk} \\ 
DAMTP,\\
Centre for Mathematical Sciences,\\
University of Cambridge, \\
Wilberforce Rd.,\\
Cambridge CB3 0WA, UK}
\date{\today}
\maketitle

\begin{abstract}
We consider the stability of spatially homogeneous
plane-wave spacetimes. We carry out a full analysis for plane-wave
spacetimes in $(4+1)$ dimensions, and find there are two cases to
consider; what we call non-exceptional and exceptional. In the
non-exceptional case the plane waves are stable to (spatially homogeneous) vacuum
perturbations as well as a restricted set of matter perturbations. In
the exceptional case we always find an instability. Also we consider the Milne universe in
arbitrary dimensions and find it is also stable provided the strong
energy condition is satisfied. This implies that there exists an open
set of stable plane-wave solutions in arbitrary dimensions.
\end{abstract}

\section{Introduction}
Finding solutions to Einstein's equations is a problem of great
mathematical and physical interest. 
However, due to
the fairly complicated nature of the Einstein equations, we usually
impose symmetries in order to make the field equations more
tractable. Some of the most successful schemes of symmetry reduction
are the Bianchi models in (3+1)-dimensional cosmology
\cite{bianchi,Taub,EM,Ellis}. For these models there exists a simply
transitive group acting on the spatial hypersurfaces; hence, they are
spatially homogeneous. For the spatially homogeneous models, the
equations of motion reduce to a simple set of ordinary differential
equations which makes it possible to study the evolution of these
models in great detail. In particular, the asymptotic behaviour of all 
non-tilted Bianchi models with a perfect fluid have been determined
\cite{Ellis,BS1,BS,HHTW,HHWV19}. Motivated by the success of the
$(3+1)$-dimensional Bianchi models, here, we shall investigate some of
their higher-dimensional counterparts. 

In $(3+1)$-dimensional cosmology, algebraically special solutions usually
play a particular role; they are equilibrium points, or sometimes even
attractors, for more general classes of solutions. In particular the
vacuum plane-wave solutions of dimension $3+1$ are attractors for
non-tilted non-inflationary models of Bianchi class B
\cite{Ellis,BS,BS1}. Hence, general 
solutions of Einstein's equations of Bianchi class B containing
non-tilted non-inflationary perfect fluids can be approximated by
vacuum plane-wave solutions at late times. For higher-dimensional
models the situation is still unsettled as no stability analysis
within this class of models has
been done to date. 

In this paper we intend to take the first step towards this and analyse
the $(4+1)$-dimensional case. More precisely we look at spatially
homogeneous plane waves, and find that they are stable to both
perturbations in the geometry and also a restricted set of matter
perturbations (which exclude inflationary fluids), provided a
particular condition is satisfied; we call this the non-exceptional
case. In the exceptional case, there exists an instability. The
exceptional cases allows -- due to the exact vanishing of one of the
Einstein constraint equations -- for an additional shear degree of
freedom; it is this extra shear mode which turns out to be
unstable. Also, as a precursor 
to future work, we show that the Milne universe is stable to
perturbations in arbitrary dimensions provided the strong energy
condition is satisfied. In particular this implies that in an
arbitrary dimension, there exists an open set of stable plane-waves.

Recently, of course, plane waves (and pp-waves) have featured heavily
in the String Theory literature \cite{BFHP,Metsaev,BMN}. Field theories in
curved backgrounds usually assume that the back-reactions on the
geometry can be neglected. However, the existence of an unstable mode
would ultimately lead to the approximation being invalid.   Therefore
knowledge of the stability 
of such spacetimes in higher dimensions is crucial to such
investigations, for studying quantum theories in such backgrounds
inevitably leads to 
issues of back-reaction. We refer the reader to the works \cite{BGS,MPZ}
where a different approach to analysing stability of plane-wave
backgrounds, in a supergravity context, is taken.

The outline of the paper is as follows. In section 2 we write down the
relevant equations of motion for spatially homogeneous models. In
section 3 we specialise to plane waves. Section 4 concerns the general
stability analysis in 4+1 dimensions. In section 5 we consider matter
perturbations, and in section 6 we analyse the stability of the
$(n+1)$-dimensional Milne universe. Finally we provide a brief
discussion.  

\section{Equations of motion and constraints in $n+1$ dimensions for
spatially homogeneous spacetimes}

In this section we recall some well known~\cite{EM,wald} results on spatially
homogeneous spacetimes in $n+1$ dimensions. Such spacetimes are
defined as ones possessing a time-like vector $n^{\mu}$ hypersurface
orthogonal, such that there exists a Lie group $G$ acting simply
transitively on the hypersurfaces. Note that this implies that the
spacetime is of the form $\mathbb{R}\times \Sigma$ with metric
$g=-dt^2 +h(t)$, where $n_{\mu}= (dt)_{\mu}$ and $h(t)$ is a left-invariant
metric on $\Sigma$. For such spaces, $n^{\mu}$ is
geodesic and the extrinsic curvature on the hypersurfaces reduces to
$K_{\mu\nu}=\nabla_{\mu}n_{\nu} \equiv \theta_{\mu\nu}$. Now, we can introduce an orthonormal
frame on our spacetime, so\footnote{Here, Greek indices ($\mu,\nu,...$)
  run over the
  full spacetime manifold, $0,1,...,n$; lower case Latin indices ($a,b,...$)
  run over the spatial hypersurfaces, $1,...,n$; and upper case Latin
  indices ($A,B,...$) run over $1,...,n-1$. }

\begin{equation}
g=\eta_{\mu\nu}\omega^{\mu}\omega^{\nu} =
-dt^2 +\delta_{ab}\omega^{a}\omega^{b},
\end{equation} where we have picked
$\omega^0=dt$. The dual basis to $\{ \omega^{\mu} \}$ will be denoted
by $\{e_{\mu} \}$, and we have 

\begin{equation}
[e_{\mu},e_{\nu}]=\gamma^{\rho}_{~\mu\nu}e_{\rho},
\end{equation} or
equivalently $d\omega^{\mu}=\Gamma^{\mu}_{~\nu\rho}\omega^{\nu} \wedge
\omega^{\rho}$, where
$\Gamma_{\rho\mu\nu}=-\frac{1}{2}(\gamma_{\rho\mu\nu}+\gamma_{\mu\nu\rho}-\gamma_{\nu\rho\mu})$.
Note that orthogonality implies that 
\begin{equation}\gamma^{0}_{~\mu\nu}=0.
\end{equation} The
remaining structure constants are
\begin{equation}
\gamma^{a}_{~0b}=-\theta^a_b+\Omega^a_{~b}
\end{equation} and $\gamma^{a}_{~bc}
\equiv C^a_{~bc}$; note $\Omega_{ab}$ is the local angular velocity of a set of
Fermi-propagated axes with respect to $\{ e_a \}$, in the rest frame
of an observer with velocity $n^{\mu}$.
The shear, $\sigma_{\mu\nu}$, is defined to be the traceless part of
$\theta_{\mu\nu}$, so $\theta_{\mu\nu}=\sigma_{\mu\nu}
+\frac{1}{n}\theta h_{\mu\nu} $. 

The structure constants $C^a_{~bc}$
can be used to define the Lie algebra of the group $G$. The Bianchi
models in $3+1$ dimensions are constructed by considering all possible
$C^a_{~bc}$. Analogously to the $3+1$ case we write these structure
constants in terms of their traceless part $D^a_{~bc}$, so
$D^a_{~ac}=0$ and 
\begin{equation}
C^a_{~bc}=D^a_{~bc} + 2\delta^a_{[b}a_{c]},
\end{equation} where
$a_b \equiv\ \frac{1}{n-1}C^a_{~ab}$. The Jacobi identity implies that
$D^a_{~bc}a_a=0$, as well as the dynamical equations,
\begin{eqnarray}
&&\frac{d}{dt}{a}_b + \frac{1}{n}\theta a_b -(\Omega^d_{\phantom{d}b}-\sigma^d_b)a_d=0, \\
&&\frac{d}{dt}{D}^a_{\phantom{a}bc} + \frac{1}{n} \theta
D^a_{\phantom{a}bc} +2(\Omega^d_{\phantom{d}[b}
-\sigma^d_{[b})D^a_{\phantom{a}c]d} + (\Omega^a_{\phantom{a}d}-\sigma^a_d)D^d_{\phantom{d}bd}=0.
\end{eqnarray}
So far the only assumption is spatial homogeneity. Now we impose the
field equations. Einstein's equations, in a vacuum, are of course
${}^{(n+1)}R_{\mu\nu}=0$. These can be used in conjunction with
standard results on hypersurfaces to give us our equations of
motion. Firstly we have~\cite{wald},
\begin{eqnarray}
{}^{(n+1)}R_{\mu\nu}n^{\mu}n^{\nu} = K^2 - K_{\mu\nu}K^{\mu\nu} -
\nabla_{\mu}(n^{\mu}\nabla_{\nu}n^{\nu}) +
\nabla_{\mu}(n^{\nu}\nabla_{\nu}n^{\mu}),
\end{eqnarray}
which can be used to derive an $(n+1)$-dimensional version of Raychaudhuri's
equation, namely,
\begin{equation}
\dot{\theta} + \frac{1}{n}\theta^2 + \sigma_{\mu\nu}\sigma^{\mu\nu}=0,
\end{equation}
where $\dot{} \equiv n^{\mu}\nabla_{\mu}$. Another nice fact is~\cite{wald},
\begin{eqnarray}
{}^{(n+1)}G_{\mu\nu}n^{\mu}n^{\nu} = \frac{1}{2} \left( {}^{(n)}R + K^2 -
K_{\mu\nu}K^{\mu\nu} \right),
\end{eqnarray}
which allows us to deduce the Friedmann constraint, 

\begin{eqnarray}
\left( 1-\frac{1}{n} \right) \theta^2 = \sigma^{\mu\nu}\sigma_{\mu\nu}
- {}^{(n)}R.
\end{eqnarray}
The ${}^{(n+1)}R_{ab} = h_{a}^{\mu}h_b^{\nu}
{}^{(n+1)}R_{\mu\nu}$  components of the vacuum equations lead to the
shear equations of motion,\footnote{Note that $\dot{\sigma}_{ab}=\frac{d}{dt}\sigma_{ab}+2\sigma^c_{(a}\Omega_{b)c}$.}

\begin{eqnarray}
\dot{\sigma}_{ab} +\theta \sigma_{ab} + {}^{(n)}R_{\langle ab\rangle}=0,
\end{eqnarray}
where $A_{\langle ab\rangle}$ of a tensor $A_{ ab }$ means the
trace-free part. 
One can calculate the curvature tensors on the hypersurface in terms
of the structure constants. We get,
\begin{eqnarray}
&&{}^{(n)}R_{ab}= b_{ab} - \frac{(n-1)}{2}(D_{abd}+D_{bad})a^d -
(n-1)a^2h_{ab}, \\
&&{}^{(n)}R = b^a_{\phantom{a}a} - n(n-1)a^2, \\
&&b_{ab} \equiv -\frac{1}{4}(2D^c_{\phantom{c}ad}D_{cb}^{\phantom{cb}d}
-D_{acd}D_b^{\phantom{b}cd} +2 D^c_{\phantom{c}ad}D^{d}_{\phantom{d}bc}).
\end{eqnarray}
Finally the Gauss-Codazzi equation,

\begin{equation}
D_aK^a_b - D_b K = {}^{(n+1)}R_{\mu\nu}n^{\nu}h^{\mu}_b
\end{equation}
where $D_a$ is the metric connection of $h$, leads to
$\theta^a_bC^b_{\phantom{b}ca} - \theta^a_c C^b_{\phantom{b}ab}=0$,
which in turn leads to the constraint,

\begin{eqnarray}
n\sigma_{ac}a^a - \sigma^a_b D^b_{\phantom{b}ac}=0.
\end{eqnarray}

Now we will recast these equations in terms of expansion-normalised
variables following the $(3+1)$-dimensional analysis \cite{Ellis}. So
we define $\Sigma_{ab} \equiv \sigma_{ab}/ \theta$, $d^b_{~ac}\equiv
D^b_{~ac}/\theta$, 
$\widetilde{\Omega}_{ab}\equiv \Omega_{ab}/ \theta$, $A_b
\equiv a_b /
\theta$, ${}^{(n)}\mathcal{R}_{ab}\equiv {}^{(n)}{R}_{ab}/\theta^2$,
and introduce a new time variable, $\tau$, by $d\tau/dt \equiv \theta$. Then
using Raychaudhuri's equation, the shear equations, the Friedmann
constraint and the constraint implied by the Gauss-Codazzi equations,
we find,
\begin{eqnarray}
&&\Sigma'_{ab} = \left( \Sigma^2 - \frac{n-1}{n} \right) \Sigma_{ab} -2\Sigma^c_{(a}\widetilde{\Omega}_{b)c}-
{{}^{(n)}\mathcal{R}_{\langle ab\rangle}}, \\
&&\left( 1-\frac{1}{n} \right) = \Sigma^2 -
{{}^{(n)}\mathcal{R}}, \\
\label{constraint1} &&n\Sigma_{ac}A^a - \Sigma^a_b d^b_{\phantom{b}ac}=0, 
\end{eqnarray}
where ${}' \equiv d/d\tau$ and $\Sigma^2\equiv
\Sigma^{ab}\Sigma_{ab}$.

At this point it is important to note that our equations of motion
look like $\mathsf{X}'=\mathsf{F}(\mathsf{X})$ where $\mathsf{X}$
is a column vector containing each of the independent dynamical
variables. This is an autonomous system of differential equations,
which lends itself to a straightforward stability analysis. Suppose
$\mathsf{X}_0$ is an equilibrium point; i.e.
\[\mathsf{F}(\mathsf{X}_0)=0. \]
Consider perturbing about this,
so $\mathsf{X} = \mathsf{X}_0 + \delta\mathsf{X}$. Then one finds,
that we can rewrite the system as
\[
\delta\mathsf{X}'=D\mathsf{F}(\mathsf{X}_0)\delta\mathsf{X}+\mathcal{O}(\delta
\mathsf{X}^2), 
\] where
$D\mathsf{F}$ is the Jacobian matrix, i.e. the matrix of derivatives
$\partial \mathsf{F}_i / \partial \mathsf{X}_j$. Hence, close to the
equilibrium point, we can drop the quadratic terms in
$\delta\mathsf{X}$ so the stability depends only on the matrix
$D\mathsf{F}(\mathsf{X}_0)$. More precisely, the stability
criterion for this system  is that if the eigenvalues all have
negative real parts, then we have stability; zero eigenvalues
need more analysis and the existence of a sole eigenvalue with
positive real part implies instability.

In particular, if the equilibrium points are actually a  space
parameterised by $m$ variables (as is the case in the subsequent
analysis), then perturbing inside this space,
i.e. $\mathsf{F}(\mathsf{X}_0+\delta \mathsf{X})=0$, implies 
\[D\mathsf{F}(\mathsf{X}_0)\delta\mathsf{X}=0; \]
namely, $\delta\mathsf{X}$ is an eigenvector of
$D\mathsf{F}(\mathsf{X}_0)$ with zero eigenvalue, and thus,
$D\mathsf{F}(\mathsf{X}_0)$ will possess at least $m$ zero
eigenvalues.

\section{Homogeneous plane-waves}

Plane-waves are solutions to Einstein's equations possessing a
covariantly constant null vector.
Now, define the null directions $\eta^{\pm}= \frac{1}{\sqrt{2}}(\omega^n
\pm \omega^0)$, and similarly $e_{\pm}=\frac{1}{\sqrt{2}}(e_n \pm e_0)$, so 

\begin{equation}
g= 2\eta^+\eta^- + \delta_{AB} \omega^A \omega^B,
\end{equation}
where $A=1,2,...,n-1$. The existence of a null covariantly constant
vector implies two things: firstly that it is a
Killing vector and secondly that its metric dual $g(\xi,\cdot)$ is
closed. We can choose $\xi$ to be parallel to $e_-$ so $\xi =  fe_-$
where $f$ is a function. Choosing a left-invariant frame (which we are free to do) implies that $[\xi,e_{A}]=0$ which gives,
\begin{eqnarray}
\gamma^{a}_{\phantom{a}A0}=\gamma^{a}_{\phantom{a}A n}, \\
e_A(f)=0.
\end{eqnarray}
Secondly, the metric dual to $\xi$, i.e. $f\eta^+$ is closed so
$d(f\eta^+)=0$ which gives,
\begin{eqnarray}
&&\gamma^{0}_{\phantom{0}ab}+\gamma^{n}_{\phantom{0}ab}=0, \\
&&\implies C^{n}_{\phantom{0}ab}=0 \quad \textrm{due to orthogonality},
\end{eqnarray}
together with a condition on $f$ which we will not deal with.
The wave front is required to be flat which tells us that there is an
$(n-1)$-dimensional Abelian subalgebra and this forces,
\begin{equation}
C^A_{\phantom{A}BC}=0.
\end{equation}
This implies that $a_B=
\frac{1}{n-1}C^a_{~aB}=\frac{1}{n-1}C^n_{~nB}=0$, which means
$a_b=a\delta_{bn}$ and upon using the evolution equation for $a_b$, we
get,
\begin{equation}
\sigma^n_A=\Omega^n_{~A}.
\end{equation}
Thus we are  left with
only the following non-zero components,

\begin{eqnarray}
C^A_{\phantom{A}Bn} \equiv \Theta^A_{\phantom{A}B}, \\
\gamma^A_{\phantom{A}B0}=C^A_{\phantom{A}Bn}.
\end{eqnarray}
For every element  $\Theta$ $\in$
$\Mat(n-1,\mathbb{R})$ we have a possible plane-wave solution. However there
exists some redundancy; namely the group of rotations $O(n-1)$ will
simply redefine our orthonormal frame $\{ \omega^A \}$. Also, we
are allowed an overall dilation, since the Einstein equations can be
used to determine $\Theta^A_{\phantom{A}A}=(n-1)a$. Thus the space of plane waves is
isomorphic to $\Mat(n-1,\mathbb{R})/[ O(n-1)\times \mathbb{R}]$; note
the dimension of this space is $\frac{1}{2}n(n-1)-1$, see Appendix B. Now, the structure constants $\gamma^a_{\phantom{a}0b} =
-\theta^a_b + \Omega^a_{\phantom{a}b}$ imply the following constraint,

\begin{eqnarray}
&&\theta^A_B - \Omega^A_{\phantom{A}B} = \Theta^A_{\phantom{A}B}.
\end{eqnarray}
We will define $T_{AB}= \Theta_{\langle AB\rangle}$, i.e the traceless part of
$\Theta$. Thus we have,

\begin{equation}
\sigma_{AB} = T_{(AB)}.
\end{equation}
Finally, the expansion-normalised $T_{AB}$ will be denoted by $\tilde{T}_{AB}$.

\section{General analysis in $4+1$ dimensions}

In this section we will describe the main results of the paper. In
Appendix A the equations of motion can be found for the $4+1$ case. We
provide the corresponding plane-wave solutions and find that
these spacetimes are indeed stable to perturbations in the geometry in
the non-exceptional case, whereas unstable in the exceptional case. We
should make a remark on exactly what sort of perturbations we are
dealing with: they correspond to perturbations in a subspace of
spatially homogeneous models, namely ones for which the structure
constants are $C^A_{\phantom{A}B4} = \Theta^A_{\phantom{A}B}$ with the
rest vanishing\footnote{All spatially homogeneous cosmological models
  of dimension 4+1 are classified in \cite{sig1}. We refer to this
  work for the other possibilities for the structure constants.}. Homogeneous plane-waves certainly belong to this class
as we showed in the previous section, although it is not clear that
all perturbations should lie in this space too. For now we assume so,
however we will discuss this issue later. 

It should also be emphasised that when we refer to stability we mean in the
technical sense; i.e. the perturbations are bounded, and not in the
asymptotic sense; i.e. all perturbations tend to zero at late times.  

\subsection{Parameterisation of independent variables}
First of all we note that the constraint (\ref{constraint1}) leads to
$\Sigma_{A4}=0$ in what we call the non-exceptional case (see Appendix
A). 
We will use our freedom to rotate the orthonormal frame to put
$T_{AB}$ in upper triangular form; note this can always be done if it
has real eigenvalues using an orthogonal matrix. The complex
eigenvalue case can also be done, but this time one needs a unitary
matrix (see Appendix B). We parameterise the variables by,

\begin{eqnarray}
&&\Sigma_{AB} = \left( \begin{array}{ccc} \mathcal{S}-2\Sigma_1 &
\Sigma_{12} & \Sigma_{13} \\ \Sigma_{12} &  \mathcal{S}+\Sigma_1
+\sqrt{3} \Sigma_2 & \Sigma_{23} \\ \Sigma_{13} & \Sigma_{23} &
\mathcal{S} +\Sigma_1 -\sqrt{3} \Sigma_2 \end{array} \right) \\
&&\tau_{AB} = \left( \begin{array}{ccc} -2\tau_1 &
\tau_{12} & \tau_{13} \\ 0  & \tau_1
+\sqrt{3} \tau_2 & \tau_{23} \\ 0 & 0  &
\tau_1 -\sqrt{3} \tau_2 \end{array} \right),
\end{eqnarray}
where $\tau_{AB} \equiv \tilde{T}_{AB}/A$. Note that 
\begin{equation}
\Sigma^2=12\mathcal{S}^2+6(\Sigma_1^2+\Sigma_2^2)+2(\Sigma_{12}^2+\Sigma_{13}^2+\Sigma_{23}^2).
\end{equation}
The independent variables left after
elimination of the constraints (see Appendix A) are $\Sigma_{AB}$ (for
$A \neq B$), $\Sigma_1$, $\Sigma_2$ and
$\tau_{AB}$ (for $A < B$), $\tau_1$, $\tau_2$; the constraints are used to
eliminate $A$ and $\mathcal{S}$. Thus we have 10 independent variables.
Note that the definition of $\tau_{AB}$ simplifies the equations of
motion, in particular the diagonal components, which correspond to the
group parameters (see Appendix B), are constants of
motion. Thus we see a further simplification and hence we are left with
effectively only 8 independent variables for each group type.

\subsection{Exact solutions}
Here we write down the general plane-wave solution to the equations of
motion. These are equilibrium points to the equations of motion. We
find,
\begin{eqnarray}
&&\tau_1=p, \quad \tau_2=q, \quad \tau_{12}=Q_1, \quad \tau_{13}=Q_2,
\quad \tau_{23}=Q_3, \nonumber\\
&&A=\frac{1}{2(\rho+2)}, \quad \mathcal{S}= -\frac{A}{2}\rho, \quad
\Sigma_1=\frac{1}{2}Ap, \quad \Sigma_2=\frac{1}{2}Aq, \nonumber\\
&&\Sigma_{12}=\frac{1}{2}AQ_1,
\quad \Sigma_{13}=\frac{1}{2}AQ_2, \quad \Sigma_{23}=\frac{1}{2}AQ_3, \nonumber
\\
&&\textrm{where}\quad \rho=p^2+q^2+\frac{1}{12}(Q_1^2+Q_2^2+Q_3^2),
\end{eqnarray}
and $p,q,Q_1,Q_2,Q_3$ are constants. Their explicit metrics are given
in \cite{sigpp}, however, these are not needed in this analysis.  Our family of plane waves 
is thus parameterised by five constants. Note that this means we expect the
Jacobian matrix to have three zero eigenvalues, since we have already
discarded the variables $\tau_1$ and $\tau_2$ on account of them being
constants of motion. These three eigenvalues correspond to
eigenvectors tangential to the set of exact plane-wave solutions. 

\subsection{Stability}
The equations of motion form an autonomous system of first order
differential equations, and as previously explained, one simply needs
the eigenvalues of the Jacobian matrix evaluated at the point we are
perturbing about. Upon linearising our system we find the following
eigenvalues:

\begin{eqnarray}
&& \lambda_1=0, \quad \lambda_2 =0 , \quad \lambda_3 =0 \nonumber\\ \nonumber
&& \lambda_4 = -\frac{3}{2(\rho+2)}, \quad \left[ \begin{array}{c}
\lambda_5 \\ \lambda_6 \end{array} \right] =  -\frac{3 \pm
2i\sqrt{Q_1^2+Q_2^2+Q_3^2}}{2(\rho+2)}, \nonumber \\
&&\left[ \begin{array}{c}
\lambda_7 \\ \lambda_8 \end{array} \right]=-\frac{3 \pm i\sqrt{Q_1^2+Q_2^2+Q_3^2}}{2(\rho+2)}.
\end{eqnarray}
Note that, ignoring the zero eigenvalues, we have 5 eigenvalues with
manifestly negative real parts (which, curiously are also all
equal). Thus in the non-exceptional case we have stability. The zero
eigenvalues were expected and correspond to moving in the space of
plane-wave solutions.

\subsection{Exceptional case}
This occurs when $T^B_{\phantom{A}A} + 4a\delta^B_A$ is not
invertible, and $a\neq 0$. Therefore the constraint leads to
$\sigma_{4B}(\tau^B_{\phantom{A}A} + 4\delta^B_A)A=0$ which implies
$2\sigma_{41}A(2-\tau_1)=0$ since $\tau$ is upper triangular. Therefore
we see the exceptional case corresponds to the particular group types
where $\tau_1=2$, which allows for
$\sigma_{41}=\Omega_{41} \neq 0$ in general. Let us restrict to $\sigma_{24}=\sigma_{34}=0$. The equation of motion
for $\Sigma_{41}$ is,

\begin{eqnarray}
\Sigma'_{41}= \left( \Sigma^2 - \frac{3}{4} -4\mathcal{S}+2\Sigma_1
\right)\Sigma_{41}.
\end{eqnarray}
Note that the exact solution is unchanged, since $\Sigma_{41}=0$
together with the exact solution in the non-exceptional case solves
all equations of motion.
If we evaluate $\Sigma^2 - \frac{3}{4} -4\mathcal{S}+2\Sigma_1$ at the exact
plane wave solution, it is easy to show that it is always greater that
$3/(\rho+2)^2$. Now, linearising the equation for
$\Sigma_{41}$ we get,

\begin{eqnarray}
\delta \Sigma'_{41} =\lambda \delta \Sigma_{41}, \\
\textrm{where} \quad \lambda >  \frac{3}{(\rho+2)^2}.
\end{eqnarray}
Therefore we have found an instability.

\section{Matter perturbations}

Having shown that the homogeneous plane waves are stable to
geometric perturbations in $(4+1)$ dimensions, it is desirable to
test their sensitivity to the insertion of matter. It proves convenient to do so in $(n+1)$ dimensions. We
begin with a stress-energy tensor for a perfect fluid,
\begin{equation}
T_{\mu\nu} = \rho n_{\mu}n_{\nu} + Ph_{\mu\nu}.
\end{equation}From the Einstein equations, one finds 
\begin{equation}
R_{\mu\nu}n^{\mu}n^{\nu} = \kappa \left({\frac{n-1}{n-2}}\rho + {\frac{n}{n-1}} P \right ).
\end{equation}
We will consider matter satisfying the barotropic equation of
state $P = (\gamma -1)\rho$. In this case the conservation of
energy implies
\begin{equation}
\dot{\rho} + \gamma \rho \theta = 0.
\end{equation} Raychaudhuri's equation now becomes
\begin{equation}\label{RAYN}
\dot{\theta} + {1 \over n}\theta^{2} + \sigma^{2} + \kappa \rho
\left({n \gamma - 2 \over n - 1} \right ) = 0.
\end{equation} We now define an expansion-normalised density,
$\Omega \equiv {\rho \over \theta^{2}}$. We can then compute its
derivative with respect to $\tau$, and using~(\ref{RAYN}) we obtain
\begin{equation}
\Omega' = \Omega \left ( -\gamma + {2 \over n} + 2 \Sigma^{2} +
{2\kappa \Omega(n\gamma - 2) \over n-1} \right).
\end{equation} To linear order, then, recalling that we are perturbing
about the vacuum, we arrive at
\begin{equation}
\delta \Omega ' =  \delta \Omega \left( -\gamma + {2 \over n} +
2\Sigma^{2}_{0} \right).
\end{equation} Here $\Sigma_{0}^{2}$
 corresponds to the vacuum plane-wave solution we are perturbing about. We immediately deduce that these spacetimes are
stable against such matter perturbations provided that 
\begin{equation}
{2 \over n} + 2 \Sigma^{2}_{0} < \gamma, 
\label{Matter}\end{equation} 
for $n=4$. This follows from the
fact that the linearised equation for $\delta\Omega$ only depends on
$\delta\Omega$ and thus the eigenvalues of the Jacobian matrix are unaltered. Given that $\Sigma^{2}$ is a
positive-definite quantity, it becomes clear demanding stability
in this model is slightly more restrictive than simply requiring
the Strong Energy Condition, namely $\gamma > {2 \over n}$, to
hold. Note that the case $\gamma=0$, which is equivalent to having a
cosmological constant, is always unstable. On the other hand, if
$\gamma=1$ as it does for dust, we can always find a stable plane wave.

\section{The Milne universe}
In this section we will use the equations set up in the first section
to analyse the stability of the Milne universe in an arbitrary dimension. This spacetime can be
defined as one with $\tilde{T}^A_{\phantom{A}B}=\Sigma_{AB}=0$; note that
this implies $A^2=1/n^2$. Explicitly the metric, in horospherical coordinates, looks like,
\begin{equation}
ds^2 =-dt^2 +t^2 \left( dz^2+e^{2z} \sum_{i=1}^{n-1} (dx^i)^2 \right).
\end{equation}
If we perturb
about this solution we get the linearised equations,

\begin{eqnarray}
&&\delta\Sigma_{AB}' = -\frac{n-1}{n} \delta\Sigma_{AB} +(n-1)A \delta
\tilde{T}_{(AB)}, \\
&&\delta \tilde{T}'^A_{\phantom{A}B} = 0,
\end{eqnarray}
where the ${}' = d/d\tau$ since we are working at linearised
level. We see that this system of autonomous equations gives rise to a
block diagonal Jacobian matrix whose eigenvalues are zero or negative. The zeros actually correspond to moving towards another
plane-wave solution. Therefore the Milne universe is stable in the
sense that the perturbations remain bounded. To see the origin of the zero
eigenvalues, write,

\begin{eqnarray}
\left( \begin{array}{c} \delta\Sigma_{AB} \\ \delta \tilde{T}_{(AB)}
\end{array} \right)' = \left( \begin{array}{cc} -\frac{n-1}{n} &
\frac{n-1}{n} \\ 0 & 0 \end{array} \right) \left( \begin{array}{c} \delta\Sigma_{AB} \\ \delta \tilde{T}_{(AB)}
\end{array} \right)
\end{eqnarray}
where we have substituted $A=1/n$ which is true for the Milne universe
($A=-1/n$ can be excluded since we can pick its orientation). Therefore, the zero eigenvalue corresponds to
$\delta\Sigma_{AB}=\delta\tilde{T}_{(AB)}$, and thus we see that we do
not move off the surface $\Sigma_{AB}=\tilde{T}_{(AB)}$ which contains
our plane waves as we saw earlier.

Interestingly, a simple argument allows us to say something about the
stability of plane-waves in arbitrary dimensions from the above
analysis. The Milne universe is the $\Sigma^2\rightarrow 0$ limit of the
set of plane-wave solutions. Observing that the Jacobian matrix is a
continuous function of the state space variables (which implies that its
eigenvalues have to be so too) we can deduce that: \emph{provided that
  eq. (\ref{Matter}) holds, there will exist
  an open set of plane-wave solutions which are stable to spatially
  homogeneous perturbations}. 

\section{Discussion}
\label{disc}
We have found that the stability analysis of vacuum spatially homogeneous
plane-waves in $4+1$ dimensions separate into two cases. For the
non-exceptional case they are stable to (spatially homogeneous) perturbations in the geometry. They are also
stable to matter perturbations provided $\gamma>1/2 +
2\Sigma^2_0$. This, of course, excludes inflationary
fluids.\footnote{The cosmological no-hair theorem tells us that in
  dimension $4+1$ for
$\gamma<1/2$ (e.g. inflationary type fluids) the late-time asymptotes
are the flat FRW universe, which is consistent with our matter
perturbation results.} The
exceptional case, however, is always unstable. This result is completely
analogous to the $(3+1)$-dimensional case where  a similar
instability for the plane-wave solution for the exceptional Bianchi
type VI$^*_{-1/9}$ \cite{HewittVI19} exists.  

It should be noted, however, if one allows for perturbations with
respect to group types then two subtleties may occur. For some group
types, even though they cannot be written as $C^A_{\phantom{A}B4} =
\Theta^A_{\phantom{A}B}$, there exist Lie algebra contractions for
which the limit is an algebra of this type \cite{Rainer}. Also, for
some special solutions an extra Killing vector may appear which may
allow for simply transitive symmetry groups other than those
considered here \cite{sigEssay}. In any case, these possibilities only arise for a set
of measure zero in the set of plane-wave solutions.

Also, we found that the Milne universe in arbitrary dimensions is stable
provided $\gamma> \frac{2}{n}$, which is the strong energy condition
for a perfect fluid. The implications of this are that in arbitrary dimensions, there always
exists a neighbourhood of stable solutions (around the Milne solution).

Of course, one can consider more general types of fluids than those
considered here; namely tilted (non-comoving) fluids, anisotropic
stresses, dissipative fluids, $p$-form fluxes, etc. For example, if one takes a tilted
perfect fluid, then generalising the arguments in \cite{BH} indicates
that non-comoving velocities may have a significant effect on the
late-time evolution of cosmological models.\footnote{A quick
  calculation shows that conservation of angular momentum in
  universes close to isotropy implies that the peculiar velocities are
  growing at late times if $\gamma>1+1/n$.} An interesting extension
of our work would be to include a systematic study of these more
general fluids.

Based on the above analysis, one is tempted to believe that all
higher-dimensional plane-wave spacetimes, except for the exceptional
models, within the class considered here,
are stable for all spatially homogeneous vacuum perturbations. We have 
already shown that some stable plane-wave spacetimes exist in any
dimension, and the stability has explicitly been shown for the (3+1)-
and (4+1)-dimensional cases. However, whether this is true in all
dimensions awaits to be seen. 

\section*{Acknowledgments}
We would like to thank John D. Barrow for useful comments and
suggestions. 
SH is funded by the Research Council of Norway and an Isaac Newton
Studentship. HKK is grateful to St. John's College, Cambridge for a
Benefactor's Scholarship. JL is funded by EPSRC. 

\appendix
\section{Equations of motion in $4+1$ dimensions}

Here follows the differential equations and constraints used in
the analysis of this paper. We will concern ourselves mainly with
the expansion-normalised variables:
\begin{eqnarray}\label{NORMALISED}
&&\Sigma_{AB}       \equiv  {\sigma_{AB} \over \theta} \nonumber
\\
&& \tilde{T}_{AB}   \equiv  {T_{AB} \over \theta} \\
&&\mathcal{S}  \equiv  {\sigma^A_{A} \over 3\theta} \nonumber \\ &&A  \equiv 
{a \over \theta} \nonumber
\end{eqnarray} Latin capital indices run from $1...3$, the
coordinates of the wave front. Further we have the following
equations
\begin{eqnarray}
&&4\sigma_{A4} a + \sigma^{4}_{B}T^{B}_{\phantom{B}A}  =  0 \nonumber \\
&&\sigma^{4}_{B} (3a \delta^{B}_{A} + \Theta^{B}_{\phantom{B}A})  =  0 \\
&&4\sigma_{44} a - \sigma^{A}_{B}T^{B}_{\phantom{B}A}  =  0. \nonumber
\end{eqnarray} In the non-exceptional case (defined as $\Theta^A_{\phantom{A}B} +3a\delta^A_B$
being invertible) we infer from the second
line above $\sigma^{4}_{A} = 0 = \Omega^{4}_{A}$.
Recall that under our parameterisation, $\sigma_{44} = -3s$. The
last of the above equations then reduces to,
using~(\ref{NORMALISED})
\begin{equation}
-12\mathcal{S} A - \Sigma_{AB}\tilde{T}^{BA} = 0.
\end{equation} 
Under the useful field redefinition, $\tau_{AB}
\equiv {\tilde{T}_{AB} \over A}$, we can solve for $\mathcal{S}$
\begin{equation}
 \mathcal{S} = - {\Sigma_{AB}\tau^{BA} \over 12}.
\end{equation} 
Although the function $\mathcal{S}$ appears in our
parameterisation of the $\Sigma^{A}_{~B}$, it will not appear on the
right hand side on account of the tracelessness of $\tau^{A}_{~B}$.

The introduction of the $\tau$ variable also
simplifies the (first order) equations of motion, which we now
produce for convenience: 

\begin{eqnarray}\label{EOM}
&&A' =  (3 \mathcal{S} + \Sigma^{2})A \nonumber \\
&&{\tau_1}'  ={\tau_2}' = 0 \nonumber \\
&& {\tau^{1}_{~2}}' =   \left(-3\Sigma_{1} - \sqrt{3}\Sigma_{2} \right ) \tau^{1}_{~2} +
2\Sigma_{12}\left(3\tau_1 +\sqrt{3} \tau_{2} \right) \nonumber
\\
 &&{\tau^{1}_{~3}}'  =  -\left( 3\Sigma_{1} -
 \sqrt{3}\Sigma_{2} \right)\tau^1_{~3} +2\Sigma_{13}\left( 3\tau_1 -\sqrt{3} \tau_{2}\right) -2\Sigma_{23}\tau^{1}_{~2} + 2\Sigma_{12}
 \tau^{2}_{~3} \nonumber \\
&& {\tau^{2}_{~3}}'  =  2\sqrt{3} \Sigma_{2}
 \tau^2_{~3} - 4\sqrt{3} \Sigma^{2}_{~3}\tau_2  \nonumber \\
&& {\Sigma_{AB}}' =   \left( \Sigma^{2} - {3 \over 4} \right)
 \Sigma_{AB} - \chi_{AB} 
\end{eqnarray}  where the symmetric matrix $\chi_{AB}$ is defined to be
\begin{eqnarray}
&&\chi_{11}  =  6A^2\tau_{1} - 2{\Sigma_{12}}^{2}- 2{\Sigma_{13}}^{2} + {5 \over
  8}A^2({\tau_{12}}^{2} + {\tau_{13}}^{2}) + {3 \over
  2}A^2({\tau_{1}}^{2} + {\tau_{2}}^{2}) \nonumber \\ 
&& \qquad + A^2{{\tau_{23}}^{2}
  \over 8} \nonumber \\
&&\chi_{12} =  -{3 A^2 \tau_{2} \over 2} -
  3\Sigma_{1}\Sigma_{12} - \sqrt{3} \Sigma_{2} \Sigma_{12} -
  2\Sigma_{13}\Sigma_{23} + {3 \over 2}A^2\tau_{1}\tau_{12}\nonumber
  \\
&& \qquad  + {\sqrt{3}
  \over 2} A^2\tau_{12} \tau_{2} + A^2{\tau_{13} \tau_{23} \over 2}
  \nonumber \\
&&\chi_{13} =  - {3 \over 2} A^2 \tau_{13} - 3\Sigma_{1} \Sigma_{13} +
  \sqrt{3} \Sigma_{13} \Sigma_{2} + {3 \over 2} A^2\tau_{1} \tau_{13} -
  {\sqrt{3} \over 2} A^2\tau_{13} \tau_{2} \nonumber \\
&&\chi_{22}  =  -3A^2(\tau_{1} + \sqrt{3}\tau_{2}) + 2{\Sigma_{12}}^{2}
  - 2{\Sigma_{23}}^{2} - {3 \over 8}A^2{\tau_{12}}^{2} + {5 \over
  8}A^2{\tau_{23}}^{2} \nonumber \\
&& \qquad + {3 \over 2}A^2({\tau_{1}}^{2} + {\tau_{2}}^{2}) +
  {1 \over 8}A^2{\tau_{13}}^{2} \nonumber \\ 
&&\chi_{23}  =  {-3 \over 2} A^2 \tau_{23} + 2\Sigma_{12} \Sigma_{13}+
  2\sqrt{3}\Sigma_{23}\Sigma_{2} - {1 \over 2}A^2\tau_{12} \tau_{13} -
  \sqrt{3} A^2\tau_{23} \tau_{2} \nonumber \\
&&\chi_{33}  =  -3A^2\tau_{1} + 3\sqrt{3} A^2 \tau_{2} +
  2{\Sigma_{13}}^{2} + 2 {\Sigma_{23}}^{2} - {3 \over
  8}A^2({\tau_{13}}^{2} + {\tau_{23}}^{2}) \nonumber \\
&& \qquad + {3 \over 2}A^2(\tau_{1}^{2} +
  \tau_{2}^{2}) + {1 \over 8}A^2{\tau_{12}}^{2}.  
\end{eqnarray}
In addition we have the Friedmann constraint, which allows us to
eliminate $A$ from the equations:
\begin{equation}
{3 \over 8} = {1 \over 2}\Sigma^{2} + A^{2}\left({\tau^{2} \over
2} + 6 \right),
\end{equation} where $\tau^{2} \equiv \tau^{(AB)} \tau_{(AB)}$.
This leads unambiguously to
\begin{equation}
A^{2} = {{3 \over 4} - \Sigma^{2} \over \tau^{2} + 12}.
\end{equation} Note that the first equation of~(\ref{EOM}) was used to derive the subsequent equations for the
$\tau^{A}_{~B}$ so it is included automatically. Thus,
$\mathcal{S}$ and $A$ have been removed and given $\tau_1$ and
$\tau_2$ (the group type) we are left with eight
dynamic variables.

\section{Some results on $n$-dimensional Lie algebras}

Two Lie algebras, given by the structure constants $C^k_{~ij}$ and
$\tilde{C}^k_{~ij}$, are isomorphic (over $\mathbb{R}$) if and only of
there exist a matrix ${\sf A}\in GL(n,\mathbb{R})$ such that 
\begin{equation} 
\tilde{C}^k_{~ij}=\left({\sf A}^{-1}\right)^k_{~l}C^l_{~pq}{\sf
A}^p_{~i}{\sf A}^q_{~j}. 
\label{LieIso}\end{equation}
We are interested in the Lie algebras where
\begin{equation}
C^A_{~BC}=0,\quad C^A_{~Bn}=\Theta^A_{~B}.
\label{def:Lie}\end{equation}
For this class of algebras, the transformation (\ref{LieIso}) can be
written in terms of $\Theta^A_{~B}$:
\begin{equation}
\tilde{\Theta}^{\tilde{A}}_{~\tilde{B}}=\left({\sf
A}^{-1}\right)^{\tilde{A}}_{~A}\Theta^A_{~B}{\sf
A}^B_{~\tilde{B}}\phi ,
\end{equation}
in some suitable new orthogonal coordinates (marked with tildes), and
$\phi$ is a non-zero constant. Hence, the Lie algebras are determined
in this case by equivalence classes of the map (in matrix form)  
\begin{equation}
{\mbold\Theta}\mapsto \phi {\sf A}^{-1}{\mbold\Theta}{\sf A}, \quad
{\sf A}\in GL(n-1,\mathbb{R}),~\phi\in \mathbb{R}-\{0\}.
\label{EqClasses}\end{equation}
The $\phi$ allows us to do a global rescaling of ${\mbold\Theta}$,
while the conjugation leaves the characteristic equation, 
\begin{equation}
\det({\mbold\Theta}-\lambda{\sf 1})=0,
\label{charEq}\end{equation}
invariant. Hence, the Lie algebras can be given in terms of the
eigenvalues of ${\mbold\Theta}$ modulo a constant rescaling. If
$\lambda_i$ are the eigenvalues, and $|\lambda_1|\neq 0$, say, then the
$(n-2)$-tuple, 
\begin{equation}
\Lambda=\left(\frac{\lambda_2}{|\lambda_1|},...,\frac{\lambda_{n-1}}{|\lambda_1|}\right),
  \qquad |\lambda_i|\geq |\lambda_{i+1}|
\end{equation}
will be invariant under the transformation (\ref{EqClasses}). This
means that different $\Lambda$ yields non-isomorphic Lie algebras. If
all eigenvalues are distinct, then $\Lambda$ will uniquely determine
the Lie algebra. However, in the degenerate cases, where two or more
eigenvalues are equal, we have to use the
\emph{Jordan canonical form} to 
determine the Lie algebra. From this one sees that in general the
space of distinct Lie algebras within the class given in eq.(\ref{def:Lie}),
is of dimension $(n-2)$.  

As an example, consider the 4-dimensional real indecomposible Lie
algebras \cite{PSW,PW,MacC}. Here, the general types are the two
two-parameter families $A^{pq}_{4,5}$ (three real eigenvalues) and
$A^{pq}_{4,6}$ (one real and two complex conjugate eigenvalues). The algebras
$A_{4,1}$-$A_{4,4}$ are the degenerate cases with non-trivial Jordan
canonical form; $A_{4,1}$ has three zero eigenvalues; $A^p_{4,2}$ has
two equal and non-zero eigenvalues; $A_{4,3}$ has two zero
eigenvalues; and  $A_{4,4}$ has three equal non-zero
eigenvalues.

In our case, ${\mbold\Theta}$ is a real matrix. This implies that the
eigenvalues $\lambda_i$ are either real, or come in complex conjugate
pairs. Due to the gauge symmetry of determining the orientation of the
frame $\{\omega^A\}$, we are allowed to perform a $O(n-1)$ rotation of
the spatial frame to simplify the matrix ${\mbold\Theta}$. 

Let us
first assume that all the eigenvalues $\lambda_i$ are real. Then for
any ${\mbold\Theta}\in \Mat(n-1,\mathbb{R})$ there exist an ${\sf R}\in
O(n-1)$ such that ${\sf R}^{-1}{\mbold\Theta}{\sf R}$ is an upper
triangular matrix. Hence, by a suitable choice of frame we can assume
that ${\mbold\Theta}$ is \emph{upper triangular} with its eigenvalues 
along the diagonal: 
\begin{equation}
{\mbold\Theta}=\begin{pmatrix}
\lambda_1 & \Theta^1_{~2} &\dotsb& \Theta^1_{~n-1} \\
0 & \lambda_2 & \dotsb & \Theta^2_{~n-1} \\
\vdots &\vdots &\ddots & \vdots \\
0 & 0 & \dotsb & \lambda_{n-1} 
\end{pmatrix}.
\label{UpperTriang}\end{equation}
Note that, as explained earlier, the ratio of the diagonal elements
determine the group type. 

If some of the eigenvalues are complex, then it is clearly not
possible to perform a real rotation so that the ${\mbold\Theta}$ is on
upper triangular form. However, using a little trick, we can still put
it into an upper triangular form. For any ${\mbold\Theta}\in \Mat(n-1,\mathbb{R})$ there exist a ${\sf U}\in
U(n-1)$ such that ${\sf U}^{\dagger}{\mbold\Theta}{\sf U}$ is an upper
triangular matrix. Using unitary matrices we can put it into the form
(\ref{UpperTriang}) with the eigenvalues along the diagonal. However,
using a unitary ``rotation'' we have to be careful about interpreting the
dynamical variables, $\sigma^A_{~B}$, $T^{A}_{~B}$ and
$\Omega^A_{~B}$. In matrix form, the unitary transformation implies
the following transformation on    $\sigma^A_{~B}$, $T^A_{~B}$:
\begin{equation}
{\mbold\sigma}\mapsto {\sf U}^{\dagger}{\mbold\sigma}{\sf U},\quad
{\bf T}\mapsto {\sf U}^{\dagger}{\bf T}{\sf U},
\end{equation}
and $\Omega^A_{~B}$ have to be reinterpreted as a
$\mathfrak{u}(n-1)$-connection:
\begin{equation}
{\mbold\Omega}\mapsto {\sf U}^{\dagger}\frac{d}{dt}{\sf U}+{\sf U}^{\dagger}{\mbold\Omega}{\sf U}.
\end{equation}
Furthermore, the equations of motion and the constraints have to be
amended similarly: 
\begin{equation}
\sigma^A_{~B}T^{B}_{~A}\mapsto {\rm Tr}({\mbold\sigma}^{\dagger}{\bf
  T}),\quad \sigma_{AB}\sigma^{AB}\mapsto {\rm Tr}({\mbold\sigma}^{\dagger}{\mbold\sigma}),
\end{equation}
etc. The equations of motion can now be solved in a similar
manner. 

For the plane-wave spacetimes, we can use this to find the dimension
of the space of plane-wave solutions. Basically, a plane-wave
spacetime is given by specifying the upper-triangular trace-free matrix
${\bf T}$. All other dynamical variables can be given, due to the
Einstein equations and constraints, and the plane-wave criterion, in
terms of ${\bf T}$. Hence, there is an $(n-2)(n+1)/2$-parameter family
of homogeneous plane-wave solutions of which $(n-2)$ parameters
specify the Lie algebra type.

\end{document}